\documentstyle[11pt]{article}
\if@twoside
\else \oddsidemargin 00pt \evensidemargin 00pt
\fi
 \newlength{\cc}\newlength{\iii}%
 \def\str{\mbox{$\scriptstyle |$}}
\def\d{{\rm d}}
\def\e{{\rm e}}
\def\i{\ifmmode{\rm i}\else\char"10\fi}
\begin{document}
\title{\bf Explicit thermostatics of certain classical one-dimensional
lattice models by harmonic analysis\thanks{
Paper presented at International Workshop on SYMMETRY METHODS IN PHYSICS,
Dubna, Russia, July 6 - 10 1993. To appear in the proceedings.}
}
\author{Georg Junker\thanks{e-mail:
junker@faupt101.physik.uni-erlangen.de}\\[2mm]
Institut f\"ur Theoretische Physik I,\\
Universit\"at Erlangen-N\"urnberg, Staudtstr.\ 7,\\
D-91058 Erlangen, Germany.}
\maketitle
\begin{abstract}
A certain class of one-dimensional classical lattice models is considered.
Using the method of abstract harmonic analysis explicit thermostatic
properties of such models are derived. In particular, we discuss the
low-temperature behavior of some of these models.
\end{abstract}
\section{A class of one-dimensional models}
In this section we will characterize a certain class of lattice models
in one dimension. This class of models has already been
considered by Romerio and Vuillermot \cite{RV74} in connection with the
transfer-matrix method. See also the book by Moraal \cite{M84} where discrete
spin models of this class are discussed.

First, we begin with the definition of the ``spin space'' denoted by $M$. We
will assume that $M$ is a homogeneous space and thus can be identified with
a group quotient $G/H$, i.e.\ $M=G/H$. Here $G$ is the transformation group
acting transitively on $M$, i.e.\ for each pair $(S,S_{0})\in M\times M$ there
exists a group element $g\in G$ such that $S=gS_{0}$
(see for example ref.\ \cite{V68} for details).
The subgroup $H\subset G$ is the stability group of some spin-direction in $M$.
We will keep this direction fixed throughout this paper and denote it by
$S_{0}$.
Hence, $hS_{0}=S_{0}$ for all $h\in H$.
For simplicity we will assume that $M$ has a finite volume and hence $G$ is a
compact group. However, all results presented below can be generalized to the
case of non-compact unimodular groups.
With the help of the unique normalized invariant Haar measure
$\d g$ on $G$ we can define a
$G$-invariant probability measure $\d S$ on the spin space $M$ \cite{M68,BR80}:
\begin{equation}
\int\limits_{M}\d S \,F(S):=\int\limits_{G}\d g\,F(gS_{0})
\end{equation}
for any integrable function $F$. The normalization and
$G$-invariance guarantee a uniform a priori probability distribution on the
spin space $M$.

In a second step, we will introduce a $G$-invariant and exchange-invariant
spin-pair interaction:
\begin{equation}
V(gS,gS')=V(S,S')=V(S',S),~~~~{\rm for~all}~g\in G.
\end{equation}
In addition we define the function
\begin{equation}
v(g):=V(S_{0},gS_{0}).
\end{equation}
Obviously, $V(S,S')=v(g^{-1}g')$
where $S=gS_{0}$ and $S'=g'S_{0}$. Furthermore, we note that $v$ is a zonal
spherical function, that is, it is invariant under left and right shifts of
the subgroup $H$:
\begin{equation}
v(hgh')=v(g)~~~~{\rm for~all}~ h,h'\in H.
\end{equation}
With the help of the spin-pair interaction (2) we can introduce a Hamiltonian
describing the interaction energy of a set of $N+1$ spins
$\{S_{1},\ldots,S_{N+1}\}$:
\begin{equation}
{\cal H}:=\sum_{j=1}^NV(S_{j},S_{j+1})=\sum_{j=1}^Nv(g_{j}^{-1}g_{j+1}),
{}~~~~ S_{j}=:g_{j}S_{0}.
\end{equation}
Viewing $S_{j}$ as the spin being attached to the $j$-th site of a
one-dimensional lattice the Hamiltonian (5) can be interpreted to
characterize a classical spin chain with nearest-neighbor coupling given
by the spin-pair interaction $V$.

The thermostatic (i.e.\ equilibrium thermodynamic) properties of such a chain
can be derived from the canonical partition function in the macroscopic limit
$N\to\infty $. The partition function, as a function of the inverse
temperature $\beta =1/k_{\rm B}T$, is defined as follows:
\begin{equation}
\begin{array}{ll}
Z_{N}(\beta )&:=\displaystyle
\int\limits_{M}\d S_{1}\cdots\int\limits_{M}\d S_{N+1}
\exp\{-\beta {\cal H}\}\\[2mm]
&\displaystyle=
\int\limits_{G}\d g_{1}\cdots\int\limits_{G}\d g_{N+1}\prod_{j=1}^N
\exp\{-\beta v(g_{j}^{-1}g_{j+1})\}.
\end{array}
\end{equation}
Note that we have used open boundary conditions for the finite chain.

In the next section we will explicitly perform the calculation of this
partition function using the method of abstract harmonic analysis and thus
obtain the free energy per spin in the macroscopic limit,
\begin{equation}
F(\beta ):=-\frac{1}{\beta }\lim_{N\to\infty }\frac{1}{N+1}\,\ln Z_{N}(\beta
),
\end{equation}
from which further thermostatic properties can be derived.
\section{Explicit thermostatic properties}
The basic fact which we will utilize for the abstract harmonic analysis on
homogeneous spaces is that the Hilbert space $L^2(M)$ decomposes, e.g.,
uniquely into
an orthogonal sum of invariant subspaces,
$L^2(M)=\bigoplus\limits_{l\in\Lambda}
H^l$. Each of these subspaces $H^l$ carries a unitary irreducible
representation which associates with each group element $g\in G$ an operator
$D^l(g)$ acting on $H^l$. The set $\Lambda $ of pairwise non-equivalent
representations consists of all so-called representations of class
one relative to $H$ \cite{V68}.
Each of these representation spaces $H^l$ contains vectors which
are invariant under arbitrary transformations $D^l(h)$, $h\in H$, of the
subgroup $H$. In the following we will make the further assumption that the
subgroup $H$ is massive, that is, there exists one and only one vector, say
$|l,0\rangle$, which is invariant in this sense \cite{V68}:
\begin{equation}
D^l(h)|l,0\rangle=|l,0\rangle,~~~~{\rm for~all}~h\in H.
\end{equation}
A sufficient condition for $H$ being massive is that $M$ is a symmetric
space \cite{M68}.
Let us denote by $\{|l,m\rangle\}$, $m=0,1,2,\ldots,d_{l}-1$, $d_{l}:=\dim
H^l$,
a complete orthonormal basis in the irreducible subspace $H^l$. Note that we
have chosen the basis such that the invariant vector (8) is one of these basis
vectors. It follows from the Peter-Weyl theorem that the matrix elements
$D^l_{00}(g):=\langle l,0|D^l(g)|l,0\rangle$, $l\in \Lambda $, form a complete
set for zonal spherical functions such as $\exp\{v(g)\}$ \cite{V68}:
\begin{equation}
\exp\{v(g)\}=\sum_{l\in\Lambda }d_{l}\,\lambda _{l}(\beta )\,D^l_{00}(g).
\end{equation}
The ``Fourier coefficients'' are given by
\begin{equation}
\lambda _{l}(\beta ):=\int\limits_{G}\d g\,\exp\{v(g)\}\,D^{l\,*}_{00}(g).
\end{equation}

Using the decomposition (9) we can put the partition function (6) into the
form
\begin{equation}
Z_{N}(\beta ):=\int\limits_{G}\d g_{1}\cdots\int\limits_{G}\d g_{N+1}
\prod_{j=1}^N\sum_{l_{j}\in\Lambda }d_{l_{j}}\,\lambda _{l_{j}}(\beta )\,
D^{l_{j}}_{00}(g_{j}^{-1}g_{j+1}).
\end{equation}
Due to the orthonormality relation
\begin{equation}
\int\limits_{G}\d g_{j}\,D^{l_{j-1}}_{00}(g_{j-1}^{-1}g_{j})
D^{l_{j}}_{00}(g_{j}^{-1}g_{j+1})=\frac{\delta _{l_{j-1}l_{j}}}{d_{l_{j}}}\,
D^{l_{j}}_{00}(g_{j-1}^{-1}g_{j+1})
\end{equation}
the calculation is straightforward. Because of the open boundary
conditions we have used, only the trivial representation, denoted by the label
$l=0\in\Lambda $, survives. The resulting partition function is explicitly
given by
\begin{equation}
Z_{N}(\beta )=\left[\lambda _{0}(\beta )\right]^N.
\end{equation}
The Fourier coefficient for the trivial representation is the average of the
statistical weight $\exp\{-\beta v(g)\}=\exp\{-\beta V(S_{0},S)\}$ with
respect to the normalized measures $\d g$ and $\d S$, respectively:
\begin{equation}
\lambda _{0}(\beta )=\int\limits_{G}\d g\, \exp\{-\beta v(g)\}
=\int\limits_{M}\d S\, \exp\{-\beta V(S_{0},S)\}.
\end{equation}

Let us note that these integral expressions can be further simplified. As
zonal spherical functions do not depend on all but only $r$ group parameters,
where $r$ is the rank of the spin space $M$ \cite{BG62},
the trivial Fourier coefficient (14) can be expressed as a $r$-fold integral.
In particular, for spin spaces of rank one all thermostatic properties
can be obtained from one single integral
expression. Important examples are so-called $n$-vector models where the
spin space is given by the unit sphere $S^{n-1}=SO(n)/SO(n-1)$ which
is of rank one. For an explicit application of the method of harmonic
analysis to such models see ref.\ \cite{JLZ93}.

The presented method of abstract harmonic analysis can be extended to
calculate also expectation values such as magnetization and two--spin
correlations \cite{JLZ93}.
\section{Discussion for low temperatures}
We are finally turning to the discussion of the low-temperature behavior of
the Fourier coefficient (14). For this we consider only the special case where
the spin space is isomorphic to a group manifold, $M\simeq G$. This case can be
embedded into the general approach of the previous section by noting that
$M=G\times G/G$. The zonal spherical functions of the product group
$G\times G$ are identical with the
characters of the representations of the group $G$ \cite{M68}. The set of
all class-one representations of $G\times G$ with respect to the subgroup $G$
can be identified with the set of all non-equivalent unitary irreducible
representations
of the group $G$. Let us stress that the function $v$ which characterizes the
spin-pair interaction (2) is now a central function, that is, it can be
decomposed into characters $\chi ^l(g):={\rm Tr\,}D^l(g)$ of unitary
irreducible representations of $G$:
\begin{equation}
v(g)=\sum_{l\in\Lambda }\,v_{l}\,\chi ^l(g)
\end{equation}
where the Fourier coefficients are given by
\begin{equation}
v_{l}:=\int\limits_{G}\d g\,v(g)\chi ^l(g^{-1}).
\end{equation}

In the following we will study the low-temperature behavior of the trivial
Fourier coefficient (14). We will first consider the case of a finite group,
hence a discrete spin model. In this case the group integrals have to be
replaced by appropriate sums:
\begin{equation}
\lambda _{0}(\beta )=\frac{1}{|G|}\,\sum_{g\in G}\exp\{-\beta v(g)\}.
\end{equation}
Here $|G|$ stands for the order of the group, that is, the number of its
elements. Similarly, we will use the notation $|{\cal P}|$ for the number of
elements
of any subset ${\cal P}\subset G$. We will make the following further
definitions:
\begin{equation}
\begin{array}{ll}
J:=\inf\limits_{g\in G}v(g),~~~~ & {\cal J}:=\{g\in G|v(g)=J\},\\[3mm]
K:=\inf\limits_{g\in G\backslash{\cal J}}v(g),~~~~ & {\cal K}:=\{g\in
G|v(g)=K\}.
\end{array}
\end{equation}
Obviously, $\Delta :=K-J>0$ is the energy gap between the ground state and
the first excited state of the chain. The numbers $|{\cal J}|$ and
$|{\cal K}|$ are the degeneracies of the
ground state and the first exited state, respectively. With these
definitions it is obvious that the leading asymptotic behavior of (17)
for large $\beta $ is given by
\begin{equation}
\lambda _{0}(\beta )\approx\frac{|{\cal J}|}{|G|}\,\e^{-\beta J}\left(1+
\frac{|{\cal K}|}{|{\cal J}|}\,\e^{-\beta \Delta }+\cdots\right).
\end{equation}
Consequently, the low-temperature behavior of the free energy, the internal
energy and the heat capacity per spin, respectively, read in the macroscopic
limit
\begin{eqnarray}
F(\beta )&=&-\frac{1}{\beta }\,\ln \lambda _{0}(\beta )=J-\frac{1}{\beta}\ln
\frac{|{\cal J}|}{|G|}-\frac{1}{\beta }\frac{|{\cal K}|}{|{\cal J}|}\,\e^{-
\beta \Delta }+\cdots,\\
E(\beta )&=&\frac{\partial}{\partial \beta }\,\left[\beta F(\beta )\right]=
J+\Delta \frac{|{\cal K}|}{|{\cal J}|}\,\e^{-\beta \Delta }+\cdots,\\
c(\beta )&=&-k_{\rm B}\beta ^2\frac{\partial E(\beta )}{\partial \beta }=
k_{\rm B}(\beta \Delta )^2\frac{|{\cal K}|}{|{\cal J}|}\,
\e^{-\beta \Delta }+\cdots.
\end{eqnarray}

In the case of a continuous group $G\simeq M$ we can find explicit
expressions if we allow only for ferromagnetic interaction, that is,
\begin{equation}
\inf\limits_{g\in G}v(g)=v(e)~~~\Longleftrightarrow
{}~~~\inf\limits_{S\in M}V(S_{0},S)=V(S_{0},S_{0})
\end{equation}
where $e\in G$ stands for the unit element of the group. Let us introduce a
para\-meterization of the group such that $g=g(\theta _{1},\cdots,\theta _{f})$
with $f:=\dim G=\dim M$. The parameters are supposed to take values in
closed intervals (for compact groups): $\theta _{a}\in [0,\theta _{a}^{\rm
max}]$. The parameterization is chosen such that the unit element is given by
$e=g(0,\ldots,0)$. Finally, if the normalized Haar measure is characterized by
a probability density $\mu (\theta _{1},\ldots,\theta _{f})$, the integral (14)
explicitly reads
\begin{equation}
\lambda _{0}(\beta )=\int\limits_{0}^{\theta _{1}^{\rm max}}\d\theta _{1}\cdots
\int\limits_{0}^{\theta _{f}^{\rm max}}\d\theta _{f}\mu (\theta _{1},\cdots,
\theta _{f})\exp\{-\beta v(g(\theta _{1},\cdots,\theta_{f}))\}.
\end{equation}
Obviously, the main contribution to this integral in the limit $\beta
\to\infty $ comes from the region with small $\theta $'s. Hence, using
Laplace's method, we expand $v$ up to second order in the group
parameters. If $\{L_{a}\}$ denotes the set of generators of $G$ we have
\begin{equation}
\begin{array}{rl}
v(g)&=\displaystyle\sum_{l\in\Lambda }v_{l}\,\chi ^l(g)=\sum_{l\in\Lambda
}v_{l}\,{\rm Tr\,}\exp\left\{\i\sum_{a=1}^f\theta _{a}L_{a}\right\}\\[2mm]
&\displaystyle\approx
\sum_{l\in\Lambda}v_{l}\,{\rm Tr\,}\left(1+\i\sum_{a=1}^f\theta _{a}L_{a}
-\frac{1}{2}\sum_{a,b=1}^f\theta _{a}\theta _{b}L_{a}L_{b}\right).
\end{array}
\end{equation}
Using the identities
\begin{equation}
{\rm Tr\,}1=d_{l},~~~~{\rm Tr\,}L_{a}=0,~~~~{\rm Tr\,}L_{a}L_{b}=\gamma
_{l}\delta _{ab},
\end{equation}
where $\gamma _{l}$ is the Dynkin index \cite{C84} of the representation
carrying label $l$, we arrive at
\begin{equation}
v(g)\approx J+\frac{k}{2}\sum_{a=1}^f\theta ^2_{a}
\end{equation}
with
\begin{equation}
J:=v(e)=\sum_{l\in\Lambda }v_{l}d_{l},~~~~k:=-\sum_{l\in\Lambda }v_{l}\gamma
_{l}.
\end{equation}
The integration may now be performed and leads to
\begin{equation}
\lambda _{0}(\beta )\approx \mu _{0}\left(\frac{2\pi }{\beta
k}\right)^{f/2}\e^{-\beta J}.
\end{equation}
Here we have assumed that $\mu _{0}:=\mu (0,\ldots,0)\neq 0$. Otherwise, the
leading non-vanishing term for small $\theta$'s has to be kept in $\mu $.
In the latter case,
we would arrive at the same result with some positive constant replacing
$\mu _{0}$.

{}From this result we obtain the leading low-temperature behavior of
the free energy, the internal energy and the heat capacity, respectively:
\begin{eqnarray}
F(\beta )&=&J-\frac{f}{2\beta}\ln \frac{2\pi}{\beta k}-
\frac{\ln\mu_{0}}{\beta}+\cdots,\\
E(\beta )&=&J+\frac{f}{2\beta}+\cdots,\\
c(\beta )&=&k_{\rm B}\frac{f}{2}+\cdots .
\end{eqnarray}
As it should be, because of the quadratic nature of (27), the low-temperature
behavior of the heat capacity is in agreement with the equipartition theorem.
This property is not restricted to the class of models discussed here, but
also holds for some anisotropic spin chains \cite{DL90}.

Finally, we note that the results for continuous spin models remain valid
for the general case $M=G/H$ if we set $f=\dim M$,
$\gamma _{l}=||L_{a}|l,0\rangle||^2$ and $\mu _{0}$ is some positive constant.
\section*{Acknowledgement}
I would like to thank the organizers for their kind invitation to
this enjoyable and inspiring conference. I am also thankful to Hajo Leschke
for his comments. Finally, support by the Heisenberg -- Landau Program is
gratefully acknowledged.

\end{document}